\documentclass{optica-article}

\journal{opticajournal} 

\articletype{Research Article}

\usepackage{nicefrac, xfrac}
\usepackage{siunitx}
\usepackage[LGRgreek]{mathastext}

\begin{document}

\title{Robust Quasi-Bound States in the continuum: Accidental and Symmetry-Protected Variants in Dielectric Metasurfaces}

\author{Hicham Mangach,\authormark{1,*} Abdelhaq Belkacem,\authormark{2} Younes Achaoui,\authormark{2} Abdenbi Bouzid,\authormark{2} Boris Gralak,\authormark{1}  and Sebastien Guenneau\authormark{3,4}}

\address{\authormark{1}CNRS, Aix Marseille Univ, Centrale Med, Institut Fresnel, Marseille, France\\
\authormark{2}Laboratory of optics, information processing, Mechanics, Energetics and Electronics, Department of Physics, Moulay Ismail University, B.P. 11201, Zitoune, Meknes, Morocco\\
\authormark{3}UMI 2004 Abraham de Moivre-CNRS, Imperial College London, SW7 2AZ, UK\\
\authormark{4}The Blackett Laboratory, Physics Department, Imperial College London, SW7 2AZ, UK}

\email{\authormark{*}hicham.mangach@fresnel.fr} 


\begin{abstract*} 
Recently, man-made dielectric materials composed of finite-sized dielectric constituents have emerged as a promising platform for quasi-bound states in the continuum (QBICs). These states allow for an extraordinary confinement of light within regions smaller than the wavelength scale. Known for their exceptional quality factors, they have become crucial assets across a diverse array of applications. Given the circumstances, there is a compelling drive to find meta-designs that can possess multiple QBICs. Here, we demonstrate the existence of two different types of QBICs in silicon-based metasurfaces: accidental QBIC and symmetry-protected QBIC. The accidental QBIC evinces notable resilience to variations in geometrical parameters and symmetry, underscoring its capacity to adeptly navigate manufacturing tolerances while consistently upholding a distinguished quality factor of $10^5$. Conversely, the symmetry-protected QBIC inherently correlates with the disruption of unit cell symmetry. As a result, a phase delay yields an efficient channel for substantial energy transference to the continuum, endowing this variant with an exceedingly high quality factor, approaching $10^8$. Moreover, the manifestation of these QBICs stems from the intricate interplay among out-of-plane electric and magnetic dipoles, alongside in-plane quadrupoles exhibiting odd parities.

\end{abstract*}

\section{Introduction}\label{sec1}

Considerable focus has been dedicated to the exploration of unique light phenomena known as bound states in the continuum (BIC) \protect\cite{kang2023applications}. Mostly observed in artificially designed structures \protect\cite{weimann2017topologically,gomis2017anisotropy,koshelev2020subwavelength}. These states possess an exceptional ability to focus and confine the intensity of electromagnetic fields within extremely small dimensions, rendering them highly appealing and significant in scientific investigations \cite{stillinger1975bound, marinica2008bound,hsu2016bound,azzam2021photonic,kupriianov2019metasurface,li2019symmetry, koshelev2018asymmetric}. The effective confinement of light with minimum absorption is a key goal in laser technology \cite{yu2021ultra}. Resonant cavities, such as photonic crystals (PhCs) with highly reflective Bragg mirrors, are commonly used for amplifying radiation \cite{cerjan2021observation, kodigala2017lasing,song2020coexistence}. Despite the use of PhC cavities, the laser's capacity to produce a single light and maintain phase coherence at certain distances remains limited \cite{schawlow1958infrared,henry1982theory}. In spectroscopy and other applications, the slender peaks found in BICs are crucial for achieving selective spectral purity and long-distance light coherence \cite{young1999visible,liang2015ultralow}. Historically, the concept of BICs originates from quantum physics,\cite{von1929remarkable,friedrich1985interfering,bulgakov2011formation} where certain electrons transcend the quantum wall limit by acquiring heightened energies and becoming confined within spatial states \cite{capasso1992observation,rybin2017supercavity}. These non-radiative states occur when charges oscillate without emitting electromagnetic radiation \cite{han2021extended,li2021free}. Given the universality of Schrödinger's equation, these dark states have been extended to various wave systems, such as photonic and phononic \cite{hsu2013observation, bulgakov2008bound, molina2012surface,chen2023phonon}. Although the quality factor (QF) and lifespan of these BICs are theoretically infinite, their actual excitation remains unattainable owing to the lack of emission mechanisms \cite{taghizadeh2017quasi,kutuzova2023quality}. In such instances, it is critical to disrupt the symmetry in order to make these perfectly non-radiative states release some energy into the nearby field \cite{koshelev2018asymmetric, sadrieva2019multipolar}. Therefore, states that are not readily imperceptible manifest conspicuously and are commonly denoted as quasi-bound states within the continuum (QBICs). Typically, there are two distinct categories of QBICs that can exist. Intentional QBICs, also known as symmetry-protected QBICs (SP-QBICs), have a particular symmetry that prevents radiation from being emitted \cite{azzam2018formation}. Accidental or Friedrich-Wintgen QBICs (FW-QBICs) belong to the second category; they occur when destructive interference effectively cancels out all radiation amplitudes in a given direction. However, this latter category is not easily accessible \cite{amrani2022friedrich,kang2021merging,yang2014analytical}. In recent endeavors, there has been notable exploration of SP-QBIC, particularly within chiral dielectric metasurfaces with high refractive indices \cite{koshelev2019nonradiating, barton2020high}. This effort has led to a substantial increase in the number of effective devices for confining energy with a small surface-to-volume ratio, thereby minimizing losses within a single oscillation cycle \cite{vuckovic2002optimization,liu2023terahertz}. One possible explanation for the sudden increase in the QF may lie in the presence of the supercavity mode \cite{han2019all}. In contrast to conventional cavity design approaches, which focus on optimizing the QF by progressively tuning the geometrical aspects of the cavity \cite{deotare2009high}. For several nanoscale applications, like integrated photonics,\cite{zhong2023photonic} high-resolution chemical and biological sensors,\cite{yin2022thz,sun2024potential} and optical modulators,\cite{tan2021active} low-loss QBICs are highly sought. This preference stems from the inherent ability of dielectrics to retain energy with minimal loss compared to plasmonic or hybrid materials, which tend to dissipate energy as heat.\\
\indent In this study, we propose a chiral dielectric metasurface capable of accommodating two distinct types of QBICs, namely FW-QBIC and SP-QBIC. The proposed design entails perturbations to the horizontal symmetry plane ($\sigma_y$), resulting in a non-centrosymmetric configuration. Primarily, altering the handedness of the unit cell, i.e., reducing its symmetrical attributes, has been theoretically and experimentally established to yield SP-QBIC \cite{huang2023realizing}. One can achieve this by adjusting the aspect ratio of one of the resonators or by simultaneously rotating the resonators clockwise and counterclockwise by equivalent degrees \cite{kim2022quasi,liu2018extreme}. Given our prior understanding that disrupting symmetry leads to the emergence of SP-QBIC, our initial investigation centers on an achiral configuration that nullifies radiation within a specific spectral range to facilitate the generation of FW-QBIC. Subsequently, we meticulously introduce chirality to generate SP-QBIC. Most importantly, the identification of these QBICs depends on the particular light polarization employed. Where the induced electric and magnetic polarizations, denoted as \textbf{P} and \textbf{M}, respectively, contribute to modes that show even and odd polarities of electric field propagation along the \textit{z}-axis. In our case, we use an incident y-polarized plane wave, where the electric (magnetic) polarization primarily aligns along the long axis $l_y$ (short axis $l_x$) of the resonators.

\section{Results}\label{sec2}

\subsection{Principle of FW/SP-QBICs Engineering}\label{subsec2.1}

\begin{figure*}[!h]
\centerline{\includegraphics[width=1\textwidth]{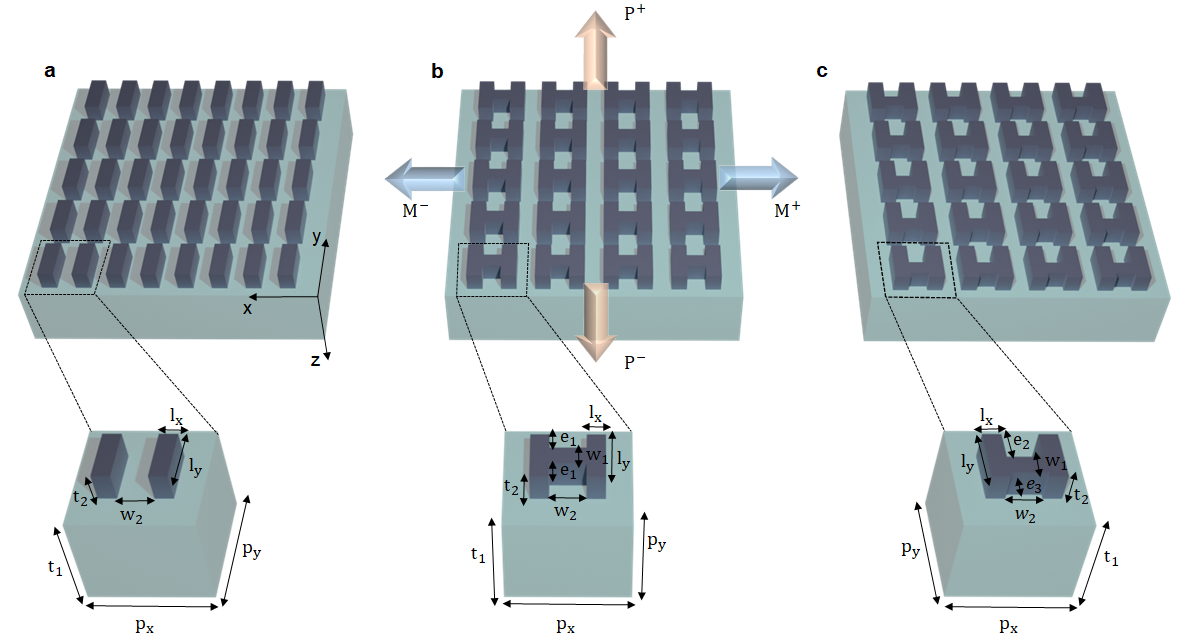}}
\caption{Metasurface prototypes: \textbf{a} In the XOY plane, the metasurface consists of pairs of identical rectangular resonators. These resonators have a center-to-center spacing of $l = w_2 + lx = 245$ \si{\nm} and show a periodicity of $p_x = p_y = 750$ \si{\nm} along both the \textit{x} and \textit{y} directions. Additionally, they possess a thickness of $t_2 = 140$ \si{\nm} and an aspect ratio of $\nicefrac{l_x}{ly} = 0.375$, with $l_y = 560$ \si{\nm}. The structural arrangement in \textbf{b} and \textbf{c} closely resembles that of \textbf{a}, with symmetrical and asymmetrical bridges connecting the resonators. The bridges have a length of $w_1 = 140$ \si{\nm}, a width of $w_2 = 70$ \si{\nm}, and maintain the same height profile as the rectangular resonators. When the symmetry is protected, we position the bridge in the center of the unit cell. However, when the symmetry is broken, the horizontal symmetry plane is disturbed by shifting the bridge by $0.0357l_y$. The rectangular resonators are placed on a silicon dioxide substrate that has a thickness of $t_{1} = 700$\si{\nm}.\label{fig1}}
\end{figure*}

For the sake of simplicity, we consider a metasurface composed of dual silicon rectangular resonators placed on a square silicon dioxide matrix with a thickness of $t_{1} = 700$ \si{\nm}. The resonators have a pitch of $750$ \si{\nm}, with both $p_{x}$ and $p_{y}$ being equal. Our analysis covers three different configurations, providing an in-depth picture of the processes that lead to two kinds of QBICs. Fig.~\ref{fig1}(\textbf{a}) displays a periodic array of pair resonators, while Fig.~\ref{fig1}(\textbf{b}) and Fig.~\ref{fig1}(\textbf{c}) emphasize a small rectangular bridge with dimensions of $w_2 = 70$ \si{\nm} and $w_1 = 140$ \si{\nm}, symmetrically and anti-symmetrically connecting the pair of resonators. The rectangular resonators are arranged with a center-to-center separation distance of $245$ \si{\nm}, while the rectangular resonators have dimensions of $l_{x} = 210$ \si{\nm} in width, $l_{y} = 560$ \si{\nm} in length, and $t_{2} = 140$ \si{\nm} in height. More details about the geometrical parameters can be found in the Fig.~\ref{fig1} caption. The choice of materials is based on the goal of minimizing energy consumption across the designated spectral range, while also considering the compatibility of the selected materials and their abundance. Furthermore, we thoroughly consider the dispersion properties of silicon and silicon dioxide, utilizing data obtained from \cite{polyanskiy2024}.

Here, we explore the propagation of a \textit{y}-polarized plane wave along the \textit{z}-axis to optically excite the system. This excitation causes charges to oscillate in accordance with the electric field orientation, resulting in an electric (magnetic) polarization. In addition to the typical dipolar charge oscillation, we can observe non-uniform charge oscillations, such as electric or magnetic quadrupoles. The intricate charge configuration in these quadrupoles contributes to the overall polarization through higher-order expansion terms. Moreover, the way dipolar and multipolar electric (magnetic) charge distributions interact with each other might provide clarity on the emerging QBICs. In this context, we quantitatively analyzed the contribution of higher-order charge interactions in the QBICs observed using the multipolar method. We are confident that this approach provides a more reliable framework for understanding the generation of these states. Although the optical properties of metasurfaces based on QBICs have often been predicted using other theoretical frameworks, like Coupled Mode Theory (CMT), these models are only effective when the symmetry of the unit cell is disturbed. To put it simply, the eigenstates of the resonators with opposite dipolar orientations cancel each other out, allowing for the detection of only the scattered energy from the background (substrate). From a multipolar analysis perspective, we take into account the contributions of both low- and higher-order electromagnetic sources in the generated QBICs. The participation of these electromagnetic sources is determined via multipole expansion formulas, which are provided as follows \cite{miroshnichenko2015nonradiating,kuznetsov2016optically}:
\begin{align}\label{Eq.01}
\textit{D}^{(e)} &= \frac{1}{\textit{i} \omega}  \int \textit{j}d^{3} \textit{r}
\end{align}

\begin{align}\label{Eq.02}
\textit{D}^{(m)} &= \frac{1}{\textit{2c}}  \int (\textit{r}\times\textit{j}) d^{3} \textit{r}
\end{align}

\begin{align}\label{Eq.03}
\textit{Q}^{(e)} &= \frac{1}{2\textit{i} \omega} \int [ (\textit{r}_{\alpha}\textit{j}_{\beta} + \textit{r}_{\beta}\textit{j}_{\alpha}) - \nicefrac{2}{3}( \textit{r}\cdot\textit{j})\delta_{\alpha,\beta} ]d^{3} \textit{r}
\end{align}

\begin{align}\label{Eq.04}
\textit{Q}^{(m)} &= \frac{1}{\textit{3c}} \int [ (\textit{r} \times \textit{j})_{\alpha} \textit{r}_{\beta} +  (\textit{r} \times \textit{j})_{\beta} \textit{r}_{\alpha}]d^{3}\textit{r}
\end{align}

The variables $\textit{D}^{(e)}$, $\textit{D}^{(m)}$, $\textit{Q}^{(e)}$, and $\textit{Q}^{(e)}$ in Eq.(\ref{Eq.01}-\ref{Eq.04}) correspond to the sources of electric and magnetic dipole and quadrupole for light, respectively. In this context, we use symbols to represent certain quantities. For example, we use $\textit{r}$ to represent the position vector, $\omega$ for the frequency, $\textit{j}$ to represent the induced current density, and $\textit{c}$ for the light velocity. Afterwards, we have used these formulas to calculate the low and higher pole participation factor of each of the observed QBICs, referred to as modes $M_1$ and $M_2$ in Fig.~\ref{fig2}(\textbf{f}). Furthermore, by using multipole formalism, each resonator is treated as a particle with a rectangular shape placed in a uniform backdrop medium. However, when the rectangular resonator is relatively small, the incident wavelength can only excites dipolar modes. In order to achieve multipolar excitation, it is necessary to use resonators of larger size. Therefore, the approximate location of these non-radiative states can be theoretically determined using a direct empirical equation ($p= 0.71 \lambda$), as demonstrated in \cite{evlyukhin2021polarization}. It is important to note that this quasi-dark state is a pure subwavelength scale phenomenon. More precisely, it can only be observed when the condition $\lambda > p$ or $kp < 2\pi$ is satisfied in certain meta-structures. Where $k$ and $p$ present the wavenumber and the unit cell period, respectively.

\subsection{Mulipoles and fields Analysis}\label{subsec2.2}

\begin{figure*}[!h]
\centerline{\includegraphics[width=\textwidth]{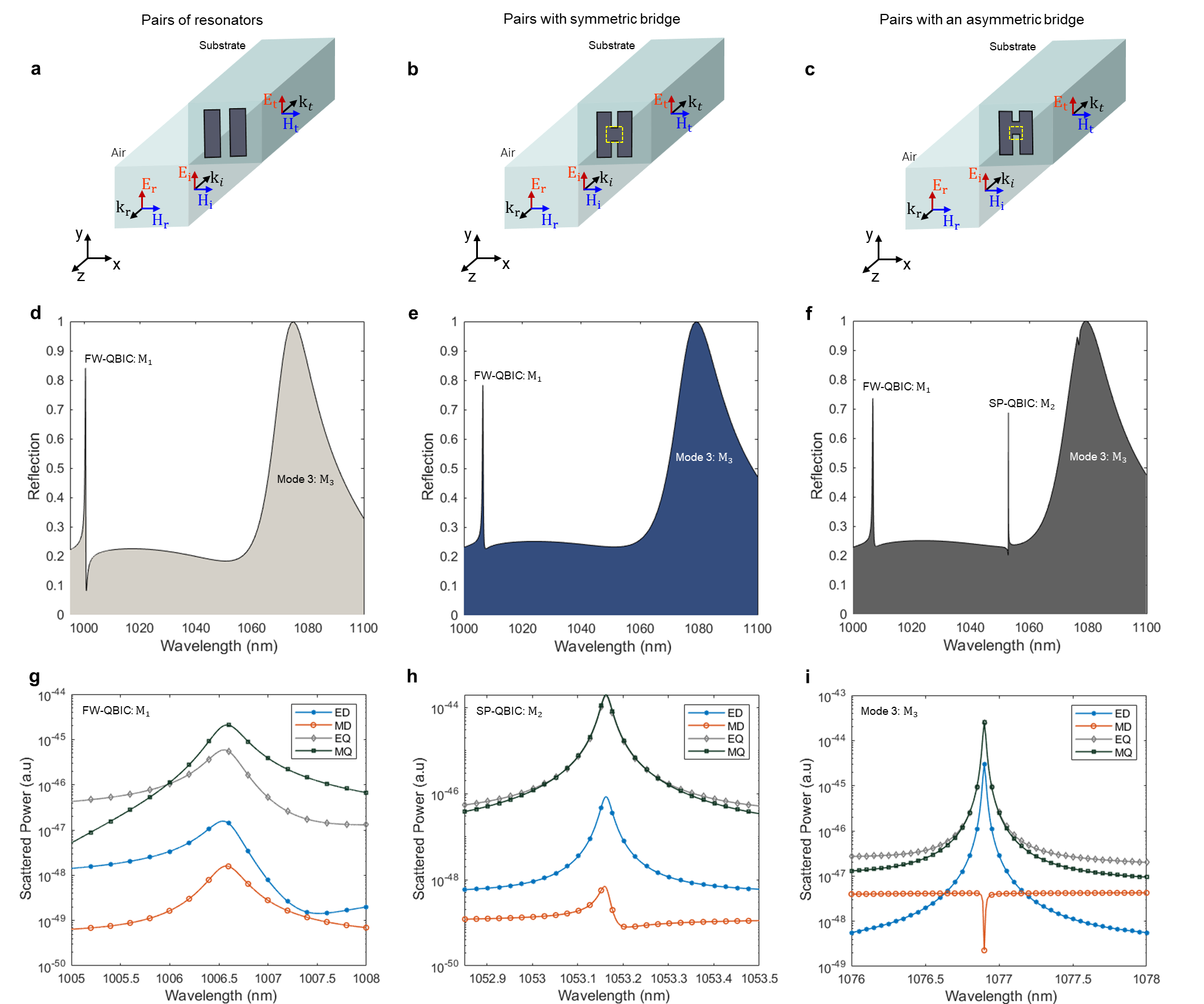}}
\caption{Illustration of exemplified QBICs: (\textbf{a}–\textbf{c}) Pairs of rectangular resonators with symmetrical properties. The FW-QBIC are depicted with and without the linking bridge. The bridge is moved along the \textit{y}-axis by a fraction equal to $\nicefrac{20}{l_y}$ in scenario (\textbf{c}), which is done on purpose to break the symmetry. (\textbf{d}-\textbf{f}) Numerical simulation of the reflection spectra of the meta-designs when excited with linear polarization parallel to the resonator's long axis. (\textbf{g}-\textbf{i}) The dominant scattered multipole power on the modes $ M_1$, $M_2$ and $M_3$ in (\textbf{f}) is elucidated through multipolar analysis.\label{fig2}}
\end{figure*}

Fig.~\ref{fig2}(\textbf{a-b}) showcases the excitation at normal incidence of an unperturbed unit cell with and without a connecting bridge. The constituent meta-atoms exhibit rotational invariance around the \textit{z}-axis, representing the $C_2$ symmetry group. Additionally, they possess two reflection symmetry planes, $\sigma_x$ and $\sigma_y$, in the YOZ and XOZ planes, respectively. Moving the connecting bridge in the \textit{y} direction alters the symmetry in  Fig.~\ref{fig2}(\textbf{c}). In this case, only the vertical symmetry $\sigma_x$ is preserved. Fig.~\ref{fig2}(\textbf{d-f}) displays the reflection spectra for each case. We have successfully proven the existence of an FW-QBIC at the wavelength of $1006.5$ \si{\nm} while maintaining the $C_2$, $\sigma_x$, and $\sigma_y$ symmetries. Upon adding the connecting bridge (refer to Fig.~\ref{fig2}(\textbf{b})), the spectral position of the FW-QBIC remains unchanged, although there is a slight change in the shape of the peak. The alteration in peak profile is expected because the silicon inclusion (resonators and bridge) occupies a larger fraction compared to Fig.~\ref{fig2}(\textbf{a}). Despite the symmetry breaking in Fig.~\ref{fig2}(\textbf{c}), the FW-QBIC remain intact. Instead, we provide evidence of a second QBIC that is protected by symmetry.
\begin{figure*}[!h]
\centerline{\includegraphics[width=\textwidth]{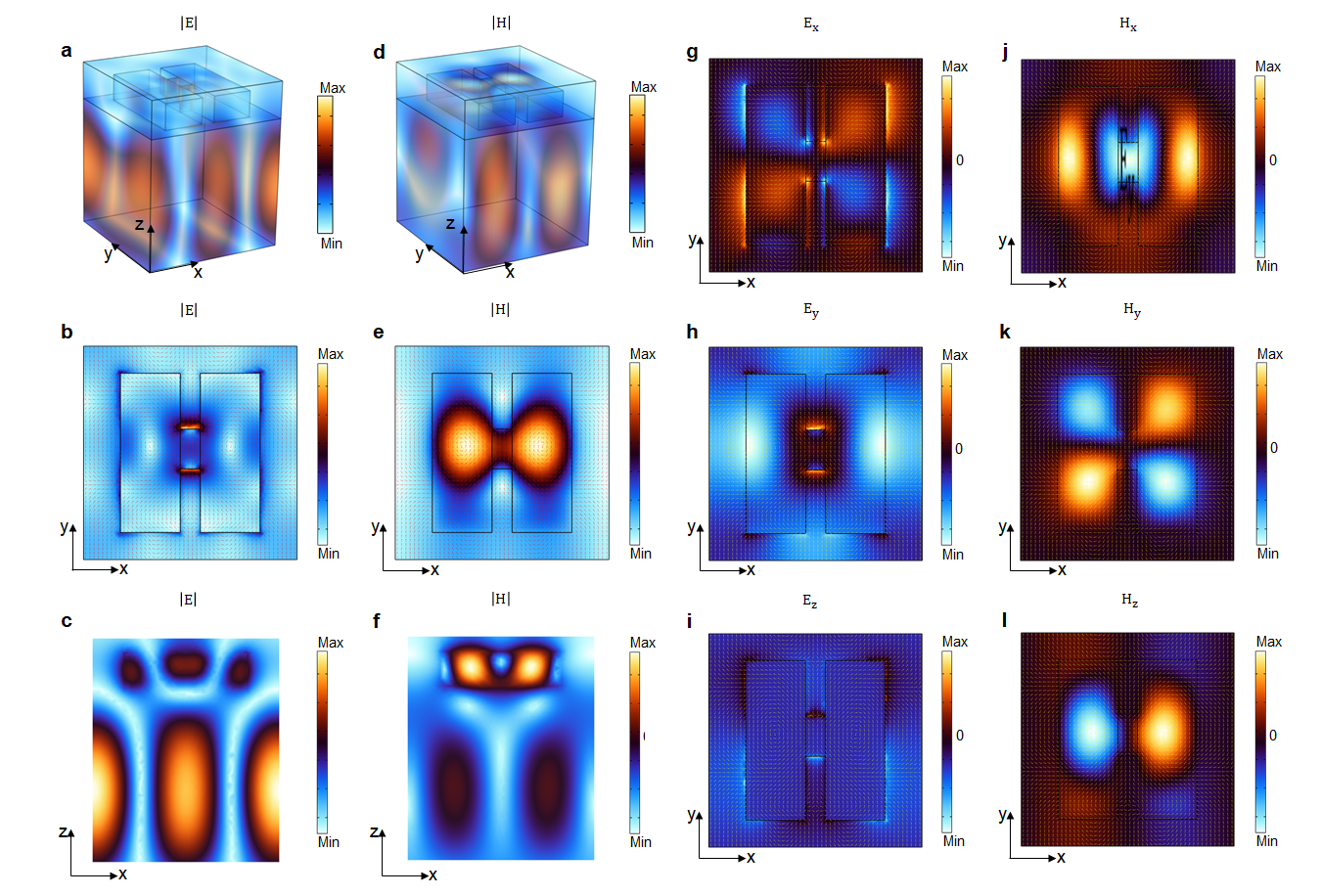}}
\caption{(\textbf{a},\textbf{d}) Three-dimensional distribution of electric and magnetic field enhancements at the FW-QBIC, located at a wavelength of $1006.5$ \si{\nm} as illustrated in Fig.~\ref{fig2}(\textbf{f}). (\textbf{b},\textbf{e}) Corresponding aerial perspective exhibiting field enhancement, observed through a top-down view employing an XY plane monitor positioned atop the silicon dioxide substrate. (\textbf{c},\textbf{f}) The field enhancement along the long axis of the resonators is profiled from a lateral perspective using an XZ plane monitor positioned centrally within the unit cell. (\textbf{g}-\textbf{l}) Visualization of surface electric and magnetic field distributions corresponding to the components $E_x$, $E_y$, $E_z$, $H_x$, $H_y$, and $H_z$, respectively. The red and yellow arrows denote the vectorial field induced by the $E_y$ field plane wave excitation. The color variation is used to improve the visual clarity compared to the backdrop surface.\label{fig3}}
\end{figure*}
Furthermore, the initial selection, denoted as M1 (FW-QBIC), exhibits strong resilience and remains unaffected by any disruptions to its symmetry. Additionally, the spectral position of the demonstrated QBICs aligns perfectly with the condition $(p = 0.71\lambda)$, as reported in previous studies. In Fig.~\ref{fig2}(\textbf{g-i}), it can be observed that the electric and magnetic quardupoles have a greater contribution compared to the dipolar ones for both FW-QBIC and SP-QBIC (modes $M_1$ and $M_2$ in Fig.~\ref{fig2}(\textbf{f})). Moreover, the SP-QBIC exhibits favorable phase alignment between low- and higher-order electric (magnetic) modes, in contrast to the FW-QBIC, which shows a phase mismatch. This discrepancy accounts for the enhanced sharpness of the second peak, $M_2$, compared to $M_1$. Subsequently, the two QBICs are stimulated by a plane wave directed from the top, with wavelengths of $1006.5$ \si{\nm} and $1053.15$ \si{\nm}, respectively. Fig.~\ref{fig3}(\textbf{a-f}) showcases the enhancement of the electric field and magnetic field at resonance.
In Fig.~\ref{fig3}(\textbf{e}), the magnetic field is concentrated in the high refractive index silicon inclusion, while in Fig.~\ref{fig3}(\textbf{b}), the electric field is concentrated at the corner point of the silicon inclusion, forming the well-known hot spot point. Fig.~\ref{fig3}(\textbf{c} and \textbf{f}) provides side views that show comparable field profile results.
The electric and magnetic field components exhibit different distributions. Specifically, the in-plane $E_x$ and $H_y$ components display a quadrupole distribution, while the out-of-plane magnetic component Hz exhibits a dipole-like distribution. It is worth noting that all the field distributions have an odd parity, except for the magnetic component $H_x$, which has an even parity, as shown in Fig.~\ref{fig3}(\textbf{g}-\textbf{l}). The FW-QBIC is then the result of two quadrupole field distributions within the plane ($E_x$ and $H_y$), as well as an out-of-the-plane dipolar magnetic component ($H_z$).

\begin{figure*}[!h]
\centerline{\includegraphics[width=\textwidth]{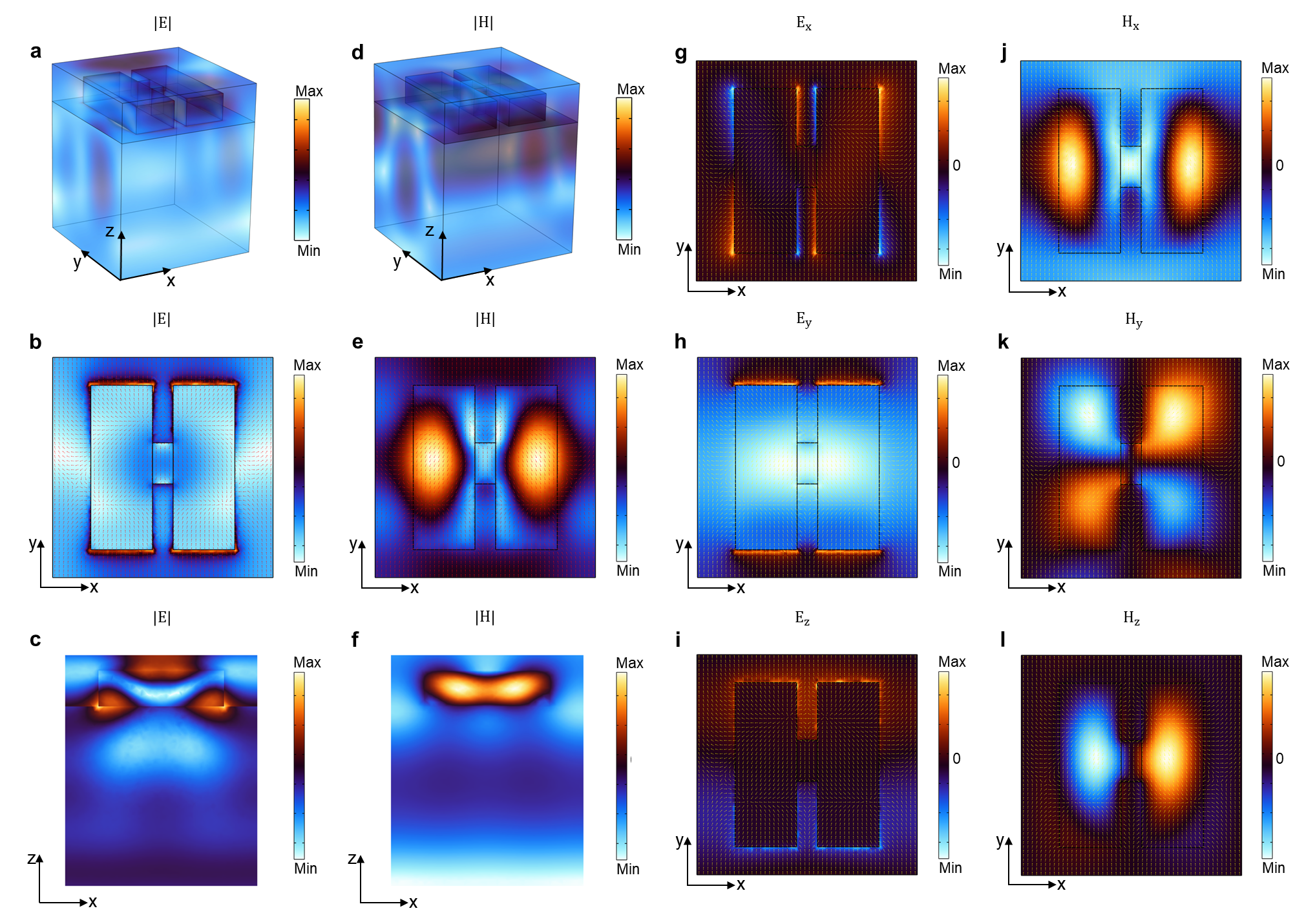}}
\caption{(\textbf{a},\textbf{d}) Visualization of the three-dimensional electric and magnetic field distributions of the SP-QBIC positioned at a wavelength of $1053.15$ \si{\nm} in Fig.~\ref{fig2}(\textbf{f}). (\textbf{b},\textbf{e}) presentation of an in-plane view, while (\textbf{c},\textbf{f}) offer side views, depicting the enhancement of surface electric and magnetic fields at resonance point M2. (\textbf{g}-\textbf{l}) The in-plane surface electric (magnetic) field components are presented as $E_x$, $E_y$, and $E_z$ ($H_x$, $H_y$, and $H_z$), respectively. The arrows in red and yellow depict the orientation of the vectorial electric field induced by the external stimulus, $E_y$. The location of the plane's monitors is the same as those used in Fig.~\ref{fig3}. \label{fig4}}
\end{figure*}

Similarly, we have performed a comparable analysis of the field distribution on the second peak (SP-QBIC) at $1053.15$ nm. Fig.~\ref{fig4}(\textbf{a}-\textbf{f}) shows a noticeable increase in the magnetic field within the silicon inclusion, while the \textit{y}-edges of the rectangular resonators exhibit a boost in the electric field. The field distribution shown in Fig.~\ref{fig4}(\textbf{g}-\textbf{l}) exhibits a similarity to the ones observed in the FW-QBIC. However, in this specific situation, there is an additional electric dipole that has an out-of-plane component, $E_z$. Moreover, when it comes to parity, the majority of the modes exhibit odd parity, with only the in-plane components $E_y$ and $H_x$ displaying even parity. It is worth mentioning that the electromagnetic field distributions in Fig.~\ref{fig3} and Fig.~\ref{fig4} match well with the multipolar results shown in Fig.~\ref{fig2}. Furthermore, the observed SP-QBIC deviates from the conventional counterpart typically observed in tilted elliptical cylinders. This deviation stems primarily from the employed polarization and the intricate interaction between dipoles and quadrupoles. Ordinarily, a wave polarized along the short axis of tilted or asymmetrical resonators is recognized to induce SP-QBICs, facilitated by the interaction between an out-of-plane magnetic dipole and an in-plane electric quadrupole \cite{moretti2024si,gao2022q}. Conversely, when the polarization aligns parallel to the resonator's long axis, this phenomenon ceases to manifest. However, in our specific investigation, the incident wave is polarized along the elongated axis of non-tilted rectangular resonators. Here, two sets of electric and magnetic dipoles and quadrupoles exist, manifesting both in-plane and out-of-plane orientations. These components act simultaneously to couple leaked energy to the continuum state.

\subsection{Angular Effect and Design Optimization}\label{subsec2.3}

\begin{figure*}[!h]
\centerline{\includegraphics[width=\textwidth]{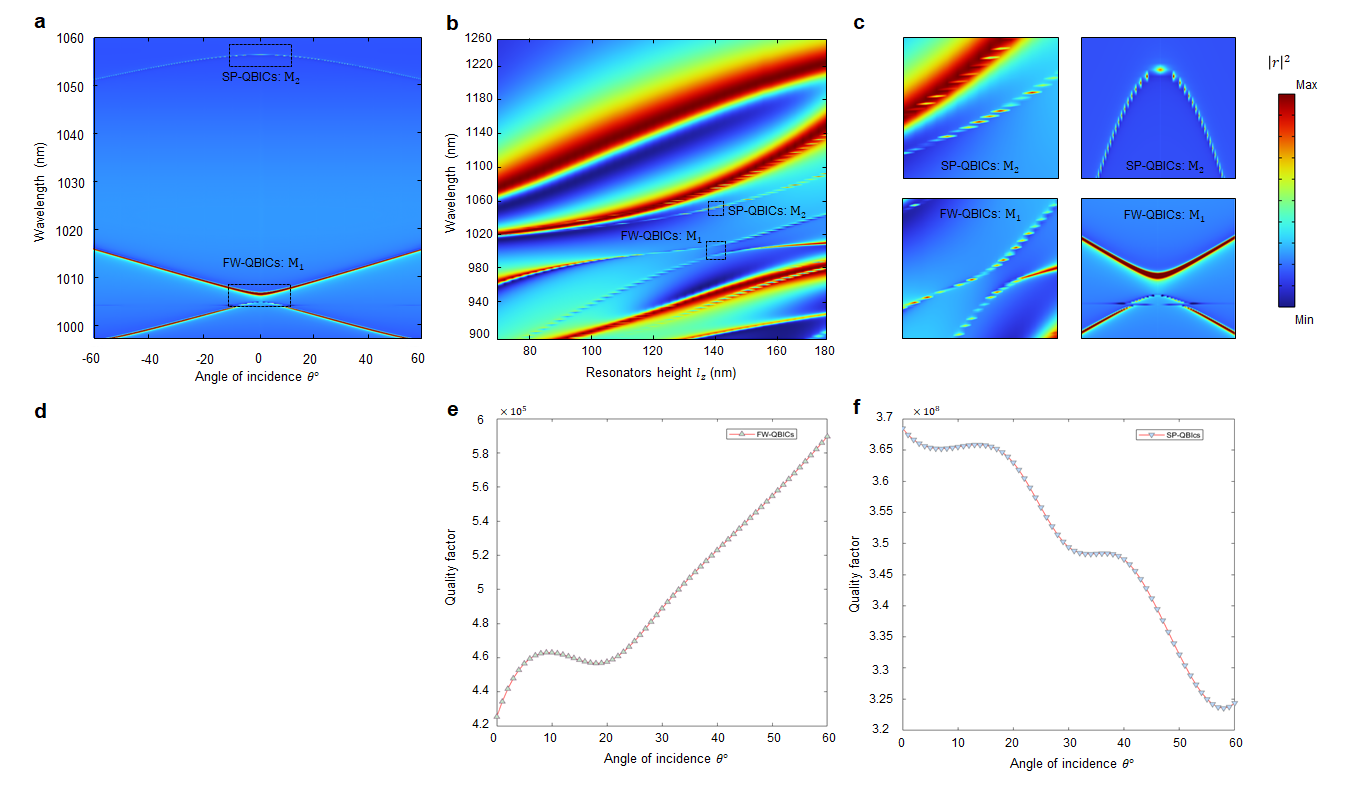}}
\caption{The field maps in (\textbf{a} and \textbf{b}) illustrate the shift in position of the QBICs due to variations in the angle of incidence and resonator height. In panel (\textbf{c}), we have included a close-up view of the critical region, which contains the singularity points in (\textbf{a} and \textbf{b}), respectively. The map in (\textbf{d}) highlights the effect of rectangular resonators on the QBICs. The panels (\textbf{e} and \textbf{f}) show the assessment of the QFs for the two QBICs, demonstrating the non-linear variations at various wavelengths. \label{fig5}}
\end{figure*}

In order to investigate the behavior and stability of the observed QBICs under different incident angles, we put the system through a plane wave with inclination angles ranging from $-60 ^{\circ}$ to $60 ^{\circ}$. As depicted in the field map of Fig.\ref{fig5}(\textbf{a}), a notable shift in the QBICs positions is evident. In the magnified view at the bottom right of Fig.\ref{fig5}(\textbf{c})), there is a noticeable energy confinement that occurs when the lower red curve becomes narrower as the incident angle increases for the FW-QBIC. Conversely, the width of the red curve for the SP-QBIC expands with increasing incidence angle. Accordingly, the FW-QBIC demonstrates its optimal QF of $5.9\times10^5$ when the incidence angle is set at 60 degrees. Conversely, the SP-QBIC exhibits its maximal QF of $3.69\times10^8$ under the condition of normal incidence. Notably, the results are markedly influenced by the resonator height, as indicated in Fig.\ref{fig5}(\textbf{b}). A singularity is observed at the inflection point of the lower curves, precisely occurring at a resonator height of $140$ \si{\nm}. Further elucidation on this phenomenon can be found in the enlarged bottom-left inset of Fig.\ref{fig5}(\textbf{c}). Additionally, multiple SP-QBICs are discernible at various resonator heights, as highlighted in the enlarged plot at the top-left corner of Fig.\ref{fig5}(\textbf{c}). This outcome underscores the pivotal role of resonator height in achieving the accidental QBIC, contrasting with the SP-QBIC's capability to exist at different resonator heights.


\section{Conclusion}
Our study presents a novel silicon metasurface capable of accommodating both accidental and symmetry-protected QBICs. The accidental QBIC showcases an impressive ability to adapt to changes in geometrical parameters, shifts in the angle of incidence, and decreases in symmetry. Conversely, there exists a significant relationship between the height of the resonators and this specific FW-QBIC. Once the height of the resonators exceeds 140 nm, this mode becomes indistinguishable. This phenomenon can only be observed when symmetry is disrupted, unlike the SP-QBIC, which relies on symmetry-breaking in the unit cell. In order to achieve our objective, we employ a bridge that connects the rectangular resonators in an asymmetrical fashion. The height of the silicon resonators does not have a significant impact on the demonstration of this SP-QBIC. It can be detected at different heights without any issue. Furthermore, a multipolar analysis is performed on each QBIC to evaluate their individual contributions, emphasizing the interplay between in-plane quadrupoles and an out-of-plane dipole as the fundamental basis for both QBICs. Furthermore, both QBICs demonstrate an impressively high QF, which makes them perfect for various applications such as lasers with super-cavity modes, high affinity spectroscopy analysis, and optical sensor platforms, amongst others.

\begin{backmatter}
\bmsection{Funding}
This work is supported by UK Research 
and Innovation (UKRI) under the UK government’s Horizon Europe funding guarantee, grant number 10033143.

\bmsection{Acknowledgments}
The section title should not follow the numbering scheme of the body of the paper. Additional information crediting individuals who contributed to the work being reported, clarifying who received funding from a particular source, or other information that does not fit the criteria for the funding block may also be included; for example, ``K. Flockhart thanks the National Science Foundation for help identifying collaborators for this work.'' 

\medskip

\bmsection{Conflicts of Interest} The authors declare no conflict of interest.

\bigskip

\bmsection{Data Availability Statement} The data that support the findings of this study are available from the corresponding author upon reasonable request.

\end{backmatter}

\bibliography{sample}






\end{document}